\documentclass[prd,aps,epsfig,showpacs,twocolumn,floats,superscriptaddress]{revtex4}
\usepackage{mathrsfs}
\usepackage{amssymb,color}
\usepackage{amssymb,graphicx,amsmath}
\begin{document}

\title{Polarization of light and the spin state of photon}
\author{Tian Guihua}
\email{tgh-2000@263.net, hua2007@126.com}
 \affiliation{School of Science, Beijing University of Posts And
Telecommunications, Beijing 100876, China. }

\begin{abstract}
The comparison of the polarization and spin of light is presented in
the paper. It is shown that it is more easier and clearer to use the
polarization of the light to explain the effect of the interaction
of light and atoms than that of spin of the light. The paper also
gives rise to the question whether or not the concept of spin for
photon have any essence for its existence.
\end{abstract}
 \maketitle

\section*{I.Introduction}
In the theory of quantum photon, the polarizations of the light are
related with the spin angular momentum $\vec{S}$ of the photon,
which can only be defined along the direction of propagation
$\vec{k}$ , the longitudinal direction. The longitudinal components
$\vec{S}\cdot\vec{ k}/|k| = m_s\hbar$ of the spin angular momentum
$\vec{S}$ can only take three values with $m_s = \pm1$ for a right-
or left-circularly polarized photon, and $m_s =0$ for a linearly
polarized one\cite{foot}-\cite{cui}.

The above statements about photon are generally accepted.
Nevertheless, the polarizations of the light is different from the
spin of photon, and sometimes their difference is so great that
substitution of the polarization of light by its spin even will
results in some error, which may be evasive in statements. We will
show this by  an example in the Zeeman effect (denoted as ZE or ZEs
in the following). In the definition of the spin for photon,the
state for the longitudinal component $m_s$ being zero is also
evasive and need some clarification. These are what this paper
intends to do.

The paper is organized as following. Section II gives brief
introduction on Zeeman effects and the explanation of the
polarization of emitted photons in ZEs by the use of the
polarization vector. In section III, ZEs is explained by use of the
conservation of the angular momentum and the concept. In this way,
it is shown they can not give a consistent answer to the
polarization concerning the $\pi$ line. Section IV give proof for
the formulae used in section III, and also shows that it is
favorable to use polarization vector to the spin of photon. Final
section ends with conclusion.

\section*{II. Zeeman effects and its correct interpretation}

ZE  is one of the crucial experiments that helped in the development
of  the quantum theory\cite{foot}. It also provides a useful tool
for examining the structure and hyperfine structure of the atoms and
molecules.   Due to the fact that magnetic fields play a more and
more important role, it can also be used to to measure the strength
and direction of magnetic fields in the process of star formation.
So, many reference books, elemental or advanced on atom theory,
contain ZE, and give explanation about it. The ZFs include the
normal and anomalous ones. Even without the quantum theory, the
spectrum and the polarization of the normal one could be easily
explained classically. However, classical theory can not give any
answer to the anomalous ZE, which needs the quantum theory,
especially the orbital and the spin angular momentums of the
electrons together to its interpretation\cite{foot} .

The peculiarity in the ZFs is the polarization of the emitted light.
In the longitudinal direction of the magnetic field $\vec{B}$, only
the $\sigma^{\pm}$ lines of the left and right circularly polarized
lights are observed, but the $\pi$ line is absent. In the transverse
direction, all three lines $\sigma^{\pm},\ \pi$ appear linearly
polarized, and $\pi$ line is polarized along the direction of the
magnetic field, while $\sigma^{\pm}$ polarized transversely to
$\vec{B}$\cite{foot}.

How to explain the polarization of the emitted light? There exist
different ways at different levels. It can be clearly stated by the
use of quantum theory of interacting atom and light system, both
atom and electromagnetical  field are quantized. Only quantum theory
of atoms can also give explanation. Of course, even classical theory
can supply the answer to those of normal ones. Most college students
contact the second explanation. However, in the second explanation
concerning the polarization of the $\pi$ line light, there is
something  misleading in some reference books, which will be picked
out in the paper\cite{yang}-\cite{cui}. The correct answer will also
be given.

As stated  before, ZFs are widely used in the modern physics and
technology, the clear and correct explanation is absolutely needed.
They also involve some fundamental concepts , like the spin and the
angular momentum for photon, etc. This is the main reason for the
paper to study this old problem.

For the sake of simplicity, we will only concern the normal ZE
regarding the explanation of the polarization of the emitted light.
As usual treatment, the dipolar polarization approximation is used
in the following. The actual calculation of the transition rate
needs the tool of time-dependent perturbation theory, however, the
Fermi's golden rule works well for our concern on the polarization
of the photon. According to the golden rule, the transition rate
$\Gamma$ is proportional to the square of the matrix element of the
perturbation of the interaction of the atom and electromagnetical
field,
\[H'=e\vec{r}\cdot\vec{E}=eE_{\vec{k},\alpha}\vec{r}\cdot\vec{e}_{\alpha},\]that is \cite{foot},
\begin{eqnarray}
\Gamma\propto
|eE_{\vec{k},\alpha}|^2|<f|\vec{r}\cdot\vec{e}_{\alpha}|i>|^2,\label{rate}
\end{eqnarray}
where $|f>,\ |i>$   are the final and initial states of the
transmitting atom. In Eq.(\ref{rate}), the emitted light's electric
filed is $\vec{E}=E_{\vec{k},\alpha}\vec{e}_{\alpha}$, and its
polarization, propagation vectors are $\vec{e}_{\alpha}$, $\vec{k}$
respectively. One could write the polarization vector in the form
$\vec{e}_{\alpha}=A^-_{\sigma}\frac{\vec{e}_x-i\vec{e}_y}{\sqrt2}+A_{\pi}\vec{e}_z+A^+_{\sigma}\frac{\vec{e}_x+i\vec{e}_y}{\sqrt2}$.
Similarly, the position vector $\vec{r}$ can be written as
\begin{eqnarray}
\vec{r}&=&r\bigg[\sin\theta\cos\phi\vec{e}_x
+\sin\theta\sin\phi\vec{e}_y+\cos\theta\vec{e}_z\bigg]\nonumber\\
&\propto
&Y_{1,-1}\frac{\vec{e}_x+i\vec{e}_y}{\sqrt2}+Y_{1,0}\vec{e}_z+Y_{1,1}\frac{\vec{e}_x-i\vec{e}_y}{\sqrt2}\label{r
vector}\end{eqnarray} So, one obtains
\begin{eqnarray}\vec{r}\cdot\vec{e}_{\alpha} &\propto &
A^-_{\sigma}Y_{1,-1}+A_{\pi}Y_{1,0}+A^+_{\sigma}Y_{1,1}\label{dot of
r e}
\end{eqnarray}
where $\vec{r},\ \vec{e}_{\alpha}$ are superposed as of three parts
corresponding to two circularly polarized vectors and a linearly
polarized one. $Y_{1,0},\ Y_{1,\pm1}$ in Eq.(\ref{r vector}) are the
spherical functions. Whenever the element
$<f|\vec{r}\cdot\vec{e}_{\alpha}|i>$ is zero, then there is no
spectral line to exist. $\sigma$-transition exists whenever $\Delta
m=m_f-m_i=\pm 1$, and $\pi$-transition hold under the condition
$\Delta m=m_f-m_i=0$. Electromagnetic radiation is a transverse
wave, so its electric field is at the right angle with its
propagation vector $\vec{k}$. When observed along the direction of
the magnetic field $\vec{B}$, there are only two $\sigma^{\pm}$
lines circularly polarized. $\pi$ line disappears due to its
polarization is the same as the direction of $\vec{B}$. On the other
hand, observed transversely, for example, along the direction of the
$x$ axis, all three lines appear with $\sigma$ ones polarized along
$y$ axis and $\pi$ line polarization vector along $z$ axis. This is
the correct explanation for the polarization of the emitted lights
in the ZFs, as already stated before. In the explanation, no
concepts such as the spin for photon, etc, have been used.

\section*{III.The contradiction encounter in Explanation of Zeeman effects with the concept of spin for photon}
Because the spherical wave functions $Y_{l,m}$ are the
eigenfunctions of the angular momentum operator $\hat{L}^2,\
\hat{L}_z$ with $l,\ m$ quantum numbers of the total and
$z$-component of the angular momentum for the atom, they extensively
suggest that the conservation of the angular momentum will give
somehow a better and simpler answer to the ZFs. As stated in the
beginning, it is a known fact a photon has energy $\hbar\omega$,
momentum $\hbar \vec{k}$ and spin angular momentum $\vec{S}$
\cite{beth}. The spin angular momentum of photon can only be defined
along the direction of propagation, and its components
$\vec{S}\cdot\vec{ k}/|k| = m_s\hbar$ with  $m_s= \pm1,\ 0$. The
photon with $m_s = \pm1$ corresponds to a right- or left-circularly
polarized light, but the state of $m_s=0$ for the photon is
amphibolous. In the following concerning the explanation of the
polarization of the $\pi$-line in the ZFs we will give precise
meaning for $m_s=0$.

 The
following is the standard explanation of ZE by the conservation of
angular momentum, which will  result in contradiction concerning the
explanation of the polarization of the $\pi$-line.

The total angular momentums are conserved for the system of the atom
and the photon. The atom's angular momentum in the
 longitudinal direction changes as $\Delta m=\pm \hbar,0$,  so, in
 the conservation of angular momentums, the emitted photon should
 have spin with $z$-component$m_s=\mp \hbar,0$. The photons of
 $z$-components $m_s=\mp \hbar$ are the $\sigma^{\pm}$ lines, which
 are circularly polarized. Of course, They appear as that  in observation
 along the direction of magnetic field $\vec{B}$. The transverse
 property of the light wave makes them as linearly polarized ones
 detected in the direction perpendicular with $\vec{B}$. As stated before, this explanation is ok
 concerning the $\sigma^{\pm}$ lines. But concerning the $\pi$ line,
 The photon has an angular momentum with its $z$-component $m_s=0$.
 It must be very easy to allure one to the conclusion that its angular
 momentum $\vec{S}$ lies in the plane perpendicular to $z$ axis\cite{yang}-\cite{cui}.
 Suppose $\vec{S}=|S|\vec{e}_x$, then the polarization vector corresponding to its components as $\pm\hbar$ must
 be the same as one of $\chi_{\pm}^x=\frac{-\vec{e}_y\mp i\vec{e}_z}{\sqrt2}$. But this viewpoint is wrong because the angular momentum is not just a common vector in quantum
 theory as it is in classical theory. Unlike the counterparts in
 classical theory, the angular momentum in quantum theory is a
 vector operator, and its components are operators not commuting
 with each other. So two or more components of the angular momentum can
 not be given simultaneously. As the $z$-component of the photon spin is
 $m_s=0$, its other components can not be determined at all.

 In the theory of quantum, the state
 for $z$-component $m_s=0$ of the
 spin angular momentum for photon is denoted as $\chi_0$, which is an unobservable state of photon whenever the photon
 is detected in the direction of $z$ axis. This just shows that
 there is no $\pi$ line to appear in the direction of the magnetic
 field $\vec{B}$. However, there is really observable effect
 concerning the state $\chi_0$. Observed perpendicular to the
 magnetic field $\vec{B}$, the state $\chi_0$ can be regarded as that of linearly polarized for
 photon. This is also shown as it is decomposed as \begin{eqnarray}
 \chi_0=\frac{1}{\sqrt2}
 (\chi_{+}^{\perp}-\chi_{-}^{\perp})\label{no commute s1}\end{eqnarray} where the superscript $\perp$ means the propagation
 direction of $\chi_{\pm}^{\perp}$ photons is at right angle with
 the $z$ axis. The above formulae will be proved later in the paper.

 For example, we choose $\chi_{\pm}^{\perp}$ as
 $\chi_{\pm}^{x}$ , the above equation becomes \begin{eqnarray}
 \chi_0=\frac{1}{\sqrt2}
 (\chi_{+}^{x}-\chi_{-}^{x})\label{no commute s}\end{eqnarray}
 As the two circularly polarized photons being real entities
 are superposed in the way of Eq.(\ref{no commute
 s}),
  they will give rise to question to the explanation concerning the polarization of
 of $\pi$-line . It means for the appearance of $\pi$-line in the transverse direction, there
 must be two photons simultaneously existing to obtain the linearly
 polarized $\pi$-line in the transverse direction, for example, the
 $x$ axis direction. Two photons appearing simultaneously will contradict with the conservation of
 energy.

 If the experiment is not about the ZFs, the use of the conservation of the sum of the angular momentum
 of the atom and the spin of the photon to explain the experiment fact can not have more errors than the one above mentioned.
  When there is a magnetic field
 $\vec{B}$ as in the ZFs, which will make the degenerate energy levels split, the above explanation is completely
 false. The photon of polarization vector $\chi_{\pm}^x=\frac{-\vec{e}_y\mp i\vec{e}_z}{\sqrt2}$
 can not exist at all.  The field $\vec{B}$ makes the energies of states with the different $m$ not be equal to each others,
 so transitions involving $\vec{e}_z,\ \vec{e}_y
 $ will have different frequencies for the corresponding photons. From
 Eq.(\ref{dot of r e}), there are three natural states for the
 photons, two circularly polarized ones ($\sigma^{\pm}$) and a linearly
 polarized one ($\pi$), and all three have different frequencies.
 Therefore, it is impossible for the photon to have such   states of
 the polarization vector $\chi_{\pm}^x=\frac{-\vec{e}_y\mp
 i\vec{e}_z}{\sqrt2}$. Because the photons of the states of
 $\chi_{\pm}^x$ can not exist at all, Eq.(\ref{no commute s}) is no
 longer having any meaning.
 The appearance of the wrong states $\chi_{\pm}^x$ for  a photon comes from the
 belief in the concept of the spin for it, which is taken
 for granted just as the well-known concepts of its energy
 $\hbar\omega$ and linear momentum $\hbar\vec{k}$.

 This in turn put the
 concept of the spin of photon in question. Though Beth shows the
 photon has spin one in 1936 \cite{beth}. Unlike the spin of
 ordinary particles, like electron, proton, etc, the spin of the
 photon must not be used without great care. It is not unreasonable
 to suggest its supersession by the simple concept of the
 polarization vector for the photon. This also is favored from the
 following consideration.

\section*{IV. Connection of polarization vector and the state for the spin}
The following will give some comments on the real meaning of the
state of photon with zero longitudinal component of its spin and the
proof of Eq.(\ref{no commute s}).

In quantum electrodynamics, the spin vector operator of the photon
is \cite{heit}, \cite{beres},
\begin{eqnarray}
                                   s_x&=& \begin{pmatrix}
                                            0 & 0 & 0 \\
                                            0 & 0 & -i \\
                                            0 & i & 0 \\
                                          \end{pmatrix},\ \
                                    s_y= \begin{pmatrix}
                                            0 & 0 & i \\
                                            0 & 0 & 0 \\
                                            -i & 0 & 0 \\
                                          \end{pmatrix}\\
                                    s_z&=& \begin{pmatrix}
                                            0 & -i & 0 \\
                                            i & 0 & 0 \\
                                            0 & 0 & 0 \\
                                          \end{pmatrix},\ \
                                          s^2= \begin{pmatrix}
                                            2 & 0 & 0 \\
                                            0 & 2 & 0 \\
                                            0 & 0 & 2 \\
                                          \end{pmatrix}
                                 \end{eqnarray}

The corresponding spin vector $\chi_{\mu}$ satisfy the following
equations \begin{eqnarray} s^2\chi_{\mu}=2\chi_{\mu},\
s_z\chi_{\mu}=\mu\chi_{\mu}\end{eqnarray} That is
\begin{eqnarray}
\chi_{0}= \ \ \ \ \begin{pmatrix}
                                                   0 \\
                                                   0 \\
                                                   1 \\
                                                 \end{pmatrix},\ \
                                                 \chi_{1}=\frac1{\sqrt2} \begin{pmatrix}
                                                   1 \\
                                                   i \\
                                                   0 \\
                                                 \end{pmatrix},\ \
                                                 \chi_{-1}=\frac1{\sqrt2} \begin{pmatrix}
                                                   1 \\
                                                   -i \\
                                                   0 \\
                                                 \end{pmatrix}
                                                 \end{eqnarray}
The physical meaning of the vectors $\chi_{1},\chi_{-1}$  is clear,
the right and the left circular polarization. Equivalently one says
that the angular spin of the photon is $\pm1$ along the direction
$z$.

The state $\chi_0$ shows that $s_z\chi_=0$. However, the state
$\chi_0$ is not defined as the state of zero of the $z$-component of
the spin for the photon, which is defined as the state of
 \begin{eqnarray}
\chi_{1}\pm \chi_{-1}
\end{eqnarray}
Actually, it is generally accepted  that the state $\chi_0$ is
physically unobservable. However, the statement must be taken with
care. It is observed in the direction of $z$ that the state $\chi_0$
is physically unobservable. Whenever we make observation in other
direction,  we could observe it. The following is the explanation.

The spin vector operator in the direction
$\vec{k}=(\sin\theta\cos\varphi,\sin\theta\sin\varphi,\cos\theta)$
is\begin{eqnarray} s_k&=&\sin\theta\cos\varphi
s_x+\sin\theta\sin\varphi s_y+\cos\theta s_z\\
s_k&=& \begin{pmatrix}
                                            0 & -i\cos\theta & i\sin\theta\sin\varphi \\
                                            i\cos\theta & 0 & -i\sin\theta\cos\varphi \\
                                            -i \sin\theta\sin\varphi& i\sin\theta\cos\varphi & 0 \\
                                          \end{pmatrix}
                                          \end{eqnarray}

The corresponding spin vector $\chi_{\mu}^k$ satisfy the following
equations \begin{eqnarray} s^2\chi_{\mu}^k=2\chi_{\mu}^k,\
s_k\chi_{\mu}^k=\mu\chi_{\mu}^k\end{eqnarray} That is
\begin{eqnarray}
\chi_{0}^k&=& \ \ \ \ \begin{pmatrix}
                                                   \sin\theta\cos\varphi \\
                                                   \sin\theta\sin\varphi \\
                                                   \cos\theta \\
                                                 \end{pmatrix}\\
                                                 \chi_{1}^k&=&\frac1{\sqrt2} \begin{pmatrix}
                                                   \sin\varphi+icos\theta\cos\varphi \\
                                                   -\cos\varphi+icos\theta\sin\varphi \\
                                                   -i\sin\theta \\
                                                 \end{pmatrix}\\
                                                 \chi_{-1}^k&=&\frac1{\sqrt2} \begin{pmatrix}
                                                   \sin\varphi-icos\theta\cos\varphi  \\
                                                   -\cos\varphi-icos\theta\sin\varphi\\
                                                   i\sin\theta \\
                                                 \end{pmatrix}
                                                 \end{eqnarray}
The spin state of the photon related to the direction $\vec{k}$ is
\begin{eqnarray}
\chi=C_0^k\chi_{0}^k+C_1^k\chi_{1}^k+C_{-1}^k\chi_{-1}^k
\end{eqnarray}
Of course, spin state of the photon related to the direction $z$ is
\begin{eqnarray}
\chi=C_0\chi_{0}+C_1\chi_{1}+C_{-1}\chi_{-1}
\end{eqnarray}
The relation between $C_i, \ C_i^k,\ i=0,\pm1$ is
\begin{eqnarray}
C_0^k&=&C_0\cos\theta+\frac{\sin\theta}{\sqrt2}(C_1e^{-i\phi}+C_{-1}e^{i\phi})\\
C_1^k&=&\frac{i\sin\theta}{\sqrt2}C_0+i\frac{1-\cos\theta}2C_1e^{-i\phi}\nonumber\\
&& -i\frac{1+\cos\theta}2C_{-1}e^{i\phi}\\
C_{-1}^k&=&-\frac{i\sin\theta}{\sqrt2}C_0+i\frac{1+\cos\theta}2C_1e^{-i\phi}\nonumber\\
&& -i\frac{1-\cos\theta}2C_{-1}e^{i\phi}
\end{eqnarray}
It is easy to see that the state $\chi=\chi_{0}$ could be written as
\begin{eqnarray}
\chi_0=\cos\theta\chi_{0}^k+\frac{i\sin\theta}{\sqrt2}
\chi_{1}^k-\frac{i\sin\theta}{\sqrt2} \chi_{-1}^k
\end{eqnarray}
From the above equation, we will obtain Eqs.(\ref{no commute s1}) ,
(\ref{no commute s}) under the assumption $\theta=\frac{\pi}2$.

Of course,  it is the state $\chi_0^k$ that is unobservable in the
direction $\vec{k}$. Hence, one still could observe the state
$\chi_0$ as $\frac{i\sin\theta}{\sqrt2} (\chi_{1}^k-\chi_{-1}^k)$,
which corresponds to the vector
\begin{eqnarray}
\frac{i\sin\theta}{\sqrt2}
(\chi_{1}^k&-&\chi_{-1}^k)=\begin{pmatrix}
                                                   -\sin\theta\cos\theta\cos\varphi \\
                                                   -\sin\theta\cos\theta\sin\varphi \\
                                                   \sin^2\theta \\
                                                 \end{pmatrix}\\
                                                 &=&\begin{pmatrix}
                                                   0 \\
                                                   0 \\
                                                   1 \\
                                                 \end{pmatrix}-\cos\theta\begin{pmatrix}
                                                   \sin\theta\cos\varphi \\
                                                   \sin\theta\sin\varphi \\
                                                   \cos\theta \\
                                                 \end{pmatrix}\\
                                                 &=&\vec{e}_z-\cos\theta
                                                 \vec{k}\label{polar0k}.
\end{eqnarray}
 From the above equation, it is easy to see that the state $\chi_0$ will be observed as  $\frac{i\sin\theta}{\sqrt2}
 (\chi_{1}^k-\chi_{-1}^k)$, linearly polarized, in the direction $\vec{k}$. Therefore, $\chi_0$ is not observed only in the direction
of $z$ axis, and will be observed  in other directions.
Eq.(\ref{polar0k}) shows that  $\frac{i\sin\theta}{\sqrt2}
 (\chi_{1}^k-\chi_{-1}^k)$
   is the  vector of the polarization  projected from the vector $\vec{e}_z$ of
 the photon in the direction orthogonal to the vector
 $\vec{k}$. So, it is reasonable to assume that $\chi_0$ state is
 one that the photon is polarized along the $z$ axis direction.
 Furthermore, as already pointed, the spin concept will give rise to
 controversies for the explanation concerning the polarization of
 $\pi$ line in ZE. Therefore,
 we consider that the polarization of light is more physically favorable than the
 concept of spin of photon.

\section*{V.Conclusion}

 Conclusion comes as following. Though  it may give somehow simpler  interpretation such as in the case of that concerning the $\sigma$-lines,
 the spin of the photon might give some wrong information as in the case related that of the $\pi$-lines in ZEs.
 All ZE experiments will be ok with the polarization vector to replace
 it. Therefore, we are more favorable in substitution of the spin of
 photon by its  polarization vector in the explanation of ZEs.

\acknowledgments The work was supported partly by the National
Natural Science of China (No. 10875018) and the Major State Basic
Research Development Program of China (973 Program: No.
2010CB923200).

\end{document}